# Data After Death: Australian User Preferences and Future Solutions to Protect Posthumous User Data

Andrew Reeves[1][0000-0001-6896-607X], Arash Shaghaghi[2][0000-0001-6630-9519], Shiri Krebs[3][0000-0003-3635-0347] and Debi Ashenden[4][0000-0003-4105-1755]

[1]Defence & Security Institute, University of Adelaide, South Australia, Australia
[2] School of Computer Science and Engineering, University of New South Wales (UNSW), Sydney, Australia.
[3]Deakin University, Geelong, Australia
[4]Institute for Cyber Security (IFCyber), University of New South Wales (UNSW), Canberra, Australia

`andrew.reeves@adelaide.edu.au`

**Abstract.**
The digital footprints of today's internet-active individuals are a testament to their lives, and have the potential become digital legacies once they pass on. Future descendants of those alive today will greatly appreciate the unprecedented insight into the lives of their long-since deceased ancestors, but this can only occur if today we have a process for data preservation and handover after death. Many prominent online platforms offer nebulous or altogether absent policies regarding posthumous data handling, and despite recent advances it is currently unclear *who* the average Australian would like their data to be managed after their death (i.e., social media platforms, a trusted individual, or another digital executor). While at present the management of deceased accounts is largely performed by the platform (e.g., Facebook), it is conceivable that many Australians may not trust such platforms to do so with integrity. This study aims to further the academic conversation around posthumous data by delving deeper into the preferences of the Australian Public regarding the management of their data after death, ultimately to inform future development of research programs and industry solutions. A survey of 1020 Australians revealed that most desired a level of control over how their data is managed after death. Australians currently prefer to entrust the management of their data to a trusted close individual or a third-party software that they can administrate themselves. As expected, social media companies ranked low regarding both trust and convenience to manage data after death. Furthermore, we found that the more active internet users have stronger desire for control over their data after death, as did people with children and those with greater levels of formal education. Unexpectedly, marital status, age, and gender did not predict preferences for posthumous data control. Future research focus should be to conceptualise and develop a third-party solution that enables these preferences to be realised. Such a solution could interface with the major online vendors (social media, cloud hosting etc.) to action the deceased's will – erasing select data, while sharing other data with selected individuals.

## 1      Introduction

In the digital age, the daily lives of individuals leave a semi-permanent presence on the internet. These digital footprints have the potential become digital memorials once they pass on (Graham et al., 2015). At present, individuals have varied levels of control over the permanence of their digital footprints. From tweets to cloud-based photo albums, online data can remain present long after the individual physically departs. Distant descendants of todays' populations will value having unprecedented insight into the lives of their long-since deceased forebears. Imagine reading what mattered most to your ancestors in a digital museum, or watching your ancestors wedding from their eyes (Lambert et al., 2018). These are possible realities for future generations, but only if today we have a process for data preservation and handover after death. As many systems currently stand, accounts become memorialised and ultimately disabled with disuse (Bellamy et al., 2013), robbing many of access to this rich record of their loved ones.

Furthermore, this data is often not easily accessible to those closest to the deceased. This raises questions and challenges in the posthumous management of digital data. Who owns this data? How should it be handled, and more importantly, who decides its fate? If managed appropriately, the data gathered today could form valuable assets for future generations. However, current industry practices mean data can be locked or deleted after the owner is deceased, leaving little for future generations to appreciate.

A significant void exists in the form of consistent procedures or policies for managing one's digital afterlife. Many prominent online platforms offer nebulous or altogether absent policies regarding posthumous data handling (Mali & Prakash, 2020). Such ambiguity not only exacerbates the challenges faced by bereaved families but also raises ethical and legal dilemmas surrounding data privacy and ownership.

Within this global landscape, Australia presents a distinct melting pot of diverse preferences. The nation's blend of multicultural values and unique social norms (Kamp et al., 2017) create a challenging task for those wishing to produce digital memorials (Allison et al., 2023). Australians, with their high internet penetration rates and strong digital engagement (Australian Communications and Media Authority, 2022), are increasingly creating vast digital legacies. Yet, the conversations and frameworks around posthumous digital data in the Australian context remain limited. Pioneering work by Arnold et al. (2017) and Allison et al. (2023) highlight Australia's distinct societal attributes and further amplify the relevance of studying the management of digital data after death, as global solutions may not wholly cater to the specific needs and concerns of the Australian people.

Therefore, this study seeks to delve deeper into the preferences of Australians regarding the posthumous management of their data and argues for the future development of a novel technical solution framework. By doing so, it aims to shed light on a domain that, while universally relevant, requires region-specific insights for comprehensive understanding and effective solutions.

## 1.1 Related Work

**Digital Legacy and Data After Death**

Recent work highlights the importance and challenges of managing digital legacies and online data after a person's death. Bellamy et al. (2013) emphasize the need for individuals to include digital registers in their wills, containing passwords and account locations, to ensure the proper distribution of online assets. Agarwal and Nath (2021) further discuss the concept of digital inheritance and the necessity for comprehensive legal arrangements to address the issue. Similarly, Mali and Prakash (2020) argue that there is a lack of uniform practices for data preservation, removal, and inheritance post-death, and that this has severe legal implications should ownership be challenged.

While many of these more recent authors have focused on the risks (particularly legal risk) pertaining to the misuse or lack of control over data after death, others have focused on the need for appropriate frameworks to enable data to be used for a broader social good. Gulotta et al. (2017) highlight the potential of digital systems to help individuals communicate aspects of their lives and be remembered after their passing. They indicate that, far from being a simple data privacy risk, embedding appropriate legal controls and technical solutions into today's technological society will allow future generations and scholars to access a rich repository of data, text, images, and videos, of people long since passed. Therefore, while the reactionary response to legal and privacy risks may be to simply prevent all future data access (through, for example, key-less encryption to all but very specific inheritors), we believe that this would be a disservice to future culture and societies. Instead, we must find ways to allow socially important data to be available and preserved for future generations, while also enabling the agency of an individual to decide how their data should be used after death. Overall, these papers underscore the need for awareness, legal frameworks, and technological solutions to effectively manage digital legacies and online data after death.

**Trust and Data Management by Online Entities**

A growing number of social media services, and other similar platforms, have begun to recognise the critical nature of posthumous data management, rolling out solutions for users to have a semblance of control over their digital footprints after death.

For instance, Meta (owner of Facebook, Instagram, and WhatsApp), introduced its "Legacy Contact" feature, allowing users to designate someone to manage parts of their account posthumously (Brubaker & Callison-Burch, 2016). This designated individual can pin a tribute post, respond to friend requests, and change the profile picture, ensuring that the deceased's profile remains a place of reverence and remembrance. However, they cannot post as the deceased, view their messages, or remove friends. If a user does not designate a Legacy Contact, the profile can be memorialized, where it remains visible, but certain features and functionalities are frozen (Brubaker & Callison-Burch, 2016). Similarly, Google has established the "Inactive Account Manager" tool (Prates et al., 2015). This allows users to determine what happens to their data across various Google services after prolonged inactivity, which might indicate the user's passing or an inability to access their account. Users can set a duration after which their account is deemed 'inactive.' Once this period elapses, Google can either notify a trusted contact or automatically delete the account and its associated data, based on the user's preferences (Prates et al., 2015).

While these solutions offer a step forward, the implementation and user-awareness levels vary. With companies like X (formerly Twitter) and others still determining how best to approach this sensitive matter, there remains a clear need for a standardised, comprehensive, and culturally-tailored approach, especially for nations with distinct values, like Australia. Furthermore, it is unclear the extent to which Australians trust social media companies to adequately manage their data after death. Recent high-profile events, including the use of Facebook data to interfere with elections, have damaged the reputation of these organisations in the eyes of the public (Hsieh-Yee, 2021). Therefore, it may be ill-advised to assume that Australians will trust such organisations to respectfully manage their data, and to follow their wishes following their death.

Recent research has borne out this lack of public trust in social media platforms. Sikun (2022) investigates the trust levels of different age groups in information released by various social media platforms, and found that trust was low across all studied demographics. Hankvist and Karlsson (2015) indicate that active users of social media perceive the trustworthiness of the platform as low due to concerns about privacy and targeted marketing, which may be a particular concern regarding the use of data for such purposes once the user is deceased. Furthermore, Kumar (2020) discusses the security risks and privacy concerns associated with sharing personal data on social networking sites, including the longer-term concerns of unilateral changes to privacy policies to which social media companies are entitled. These concerns take on an even more profound significance when considering the potential misuse of data posthumously, a time when the deceased can't voice opposition to any unsolicited use of their digital footprints.

**Demographic Variations in Data Management Preferences**

Recent papers suggest that demographic variables such as age, gender, income, and education can influence data management preferences. Phang et al. (2010) found that age, income, and education have effects on online consumers' store visit strategies. Djamasbi and Wilson (2015) highlighted the role of demographic factors in the use of specific e-health services. Morgan et al. (2015) showed that demographic differences, including race, gender, age, socio-economic status, and education, influence health information searching behaviour. Baack et al. (2000) found that demographic variables like age, education, gender, income, and length of residence in a community can indicate differences in preferences for future city services. These findings suggest that demographics play a significant role in shaping data management preferences and trust in social media platforms. Therefore, the current study explores demographic differences in preferences in the Australian population.

**The present study**

This study sought to address the following four research questions (RQ):

(RQ1) Do Australians want control of how their data should be managed after they pass on?

(RQ2) Who do Australians trust to manage their data after death? (Social media companies, digital wills, trusted individuals, third party)

(RQ3) Do these preferences and trust levels change by demographic? (Age, gender, marital status, children, religiosity, income, internet usage activity).

(RQ4) What should any new technological solution provide to meet the needs of Australians wishing to secure their data after death?

## 2 Methodology

### 2.1 Materials

We adopted a survey methodology for this study. This methodology was deemed appropriate given our objective to understand the preferences and trust levels of Australians regarding posthumous data management.

**Demographics**
The questionnaire incorporated demographic questions to capture data on age, gender, marital status, internet usage activity, and other demographics that we considered may be relevant to explaining individuals' preferences for the management of their data after death.

**Internet Usage Activity Scale**
The Internet Usage Activity Scale is a nine-item measure developed for this study. It includes items relating to common activities that people perform on the internet, such as checking emails and accessing new websites. Participants are asked to rate their agreement with each item along a 5-point Likert scale from *Daily, Often, Sometimes, Rarely, Never*. These scores were then split into three groups of approximately equal size: average internet users (defined as the middle 50% of the data, $25^{th}$ percentile to $75^{th}$ percentile), high internet users (defined as scores greater than the $75^{th}$ percentile) and low internet users (defined as scores less than the $25^{th}$ percentile), a similar process as recommended by DeCoster et al. (2011). The Cronbach alpha for the Internet Usage Activity Scale was observed as .89 in our sample.

**Data After Death Control Preferences Scale**
The DaD Control Scale is a four-item measure developed for this study. It aims to explore the extent to which individuals feel a desire to have influence over how their data is managed once they pass on. Participants are asked to rank their agreement with the items on a five-point Likert scale from Strongly Disagree to Strongly Agree. Items include *"Users should have the option to configure how their account should be managed by a service provider after they pass away"* and *"Protecting my personal information after I pass away is an important concern for me"*. The full measure is available in Table 2. The Cronbach's alpha score in our sample is in the acceptable range ($\alpha = .68$).

### 2.2 Data Collection

We recruited 1,179 Australians through the online survey platform, Qualtrics. Qualtrics uses a closed invitation-only panel recruitment method, ensuring that participants are representative and eligible for the current study. The survey participants comprised a representative sample of adult Australian internet users in June 2021. Quotas of gender, age, income, level of education, and geographical location were set in advance and controlled during the data collection to ensure a representative sampling of Australian Internet users. Participants were required to be living in Australia and be over the age of 18. Steps were taken to ensure the confidentiality of respondents' data, including minimizing unnecessary collection of identifiable details, and informed consent was obtained prior to participation. Ethical approval was granted by the University of NSW HREC approval number SEBE-2021-15.

### 2.3    Data Analysis

The collected data underwent preliminary cleaning and processing. Incomplete responses totalling 159 were removed, producing a final dataset of 1020 valid responses. Descriptive statistics, specifically histograms, pie charts, means, and standard deviations, were utilized to provide a general overview of the respondents' preferences and trust levels, and to explore RQ1 (How do Australians wish their data would be managed after they pass on?), RQ2 (Who do Australians trust to manage their data after death?), and RQ4 (What should a third-party solution provide to meet the needs of Australians wishing to secure their data after death?). Subsequently, we performed regression analyses were to identify patterns among different demographic groups and to answer RQ3.

# Results

## 2.4 Demographics

Table 1 presents a summary of the demographics of our sample. Our sample is sufficiently diverse to represent a broad spectrum of opinions within the wider Australian community.

**Table 1.** Demographic characteristics of the sample.

| Variable | n | % |
|---|---|---|
| Gender | | |
|     Female | 525 | 51.5 |
|     Male | 493 | 48.3 |
|     Other | 2 | .2 |
| Age | | |
|     18-24 | 152 | 14.9 |
|     25-34 | 202 | 19.8 |
|     35-44 | 213 | 20.9 |
|     45-54 | 130 | 12.7 |
|     55-64 | 150 | 14.7 |
|     65-74 | 118 | 11.6 |
|     75 and above | 55 | 5.4 |
| Education | | |
|     No school | 4 | .4 |
|     Some school | 110 | 10.8 |
|     Year 12 or equivalent | 300 | 29.4 |
|     Advanced diploma and/or certificate level III/IV | 310 | 30.4 |
|     Bachelor's degree | 189 | 18.5 |
|     Master's degree | 94 | 9.2 |
|     PhD or equivalent | 13 | 1.3 |
| Annual Household Income (AUD) | | |
|     0-35,000 | 168 | 16.5 |
|     35,000-100,000 | 346 | 34.9 |
|     100,001-250,000 | 455 | 44.6 |
|     250,001 or above | 41 | 4.0 |
| Australian State | | |
|     New South Wales | 317 | 31.1 |
|     Victoria | 278 | 27.3 |
|     Queensland | 212 | 20.8 |
|     South Australia | 82 | 8.0 |
|     Western Australia | 78 | 7.6 |
|     Australian Capital Territory | 26 | 2.5 |
|     Tasmania | 22 | 2.2 |
|     Northern Territory | 5 | .5 |
| Marital Status | | |
|     Never married | 237 | 23.2 |
|     De Facto | 147 | 14.4 |
|     Married | 505 | 49.5 |
|     Divorced/Separated | 88 | 8.6 |
|     Widowed | 30 | 2.9 |
|     Other | 13 | 1.3 |
| Children or Dependents | | |
|     Yes | 599 | 58.7 |
|     No | 418 | 41.0 |
|     Other | 3 | .3 |

### 2.5 Results in relation to Research Questions

We utilised a series of descriptive and inferential methods to answer the targeted research questions. The following section presents the results against each research question.

**Do Australians want control over how their data should be managed after they pass on?**
To explore the first research question, we examined the patterns of responding to the DaD Control Scale. Table 2 provides an overview of the descriptive statistics of the measure.

**Table 2.** Descriptive Statistics for the DaD Control Scale

| Item | Mean | SD |
|---|---|---|
| Users should have the option to configure how their account should be managed by a service provider after they pass away | 3.14 | 0.95 |
| Protecting my personal information after I pass away is an important concern for me. | 2.54 | 1.07 |
| When choosing online digital services, the privacy of my information after I pass away is important to me. | 2.20 | 1.18 |
| When choosing online digital services, having control over what happens to my information after I pass away is important to me. | 2.18 | 1.21 |

The following set section presents figures that further explain the preferences we observed in our sample.

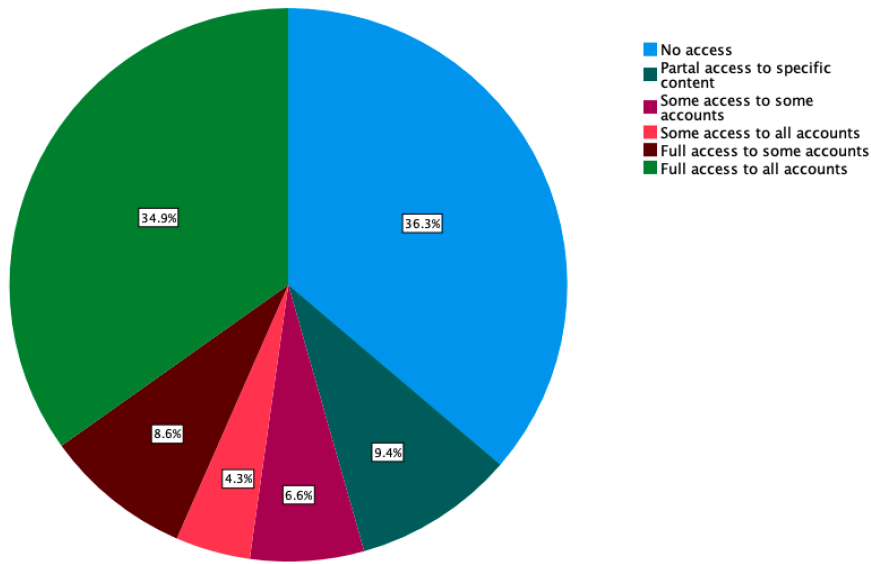

**Figure 1.** Frequency of responses to DaD Control Scale Item 1: *Would you like another person to have access to your online accounts after you pass away?*

As indicated in Figure 1, the plurality of respondents indicated that they would like another person to have full access to all their online accounts after they pass away. This was followed closely by those who preferred that no access be provided to any one person after they pass away.

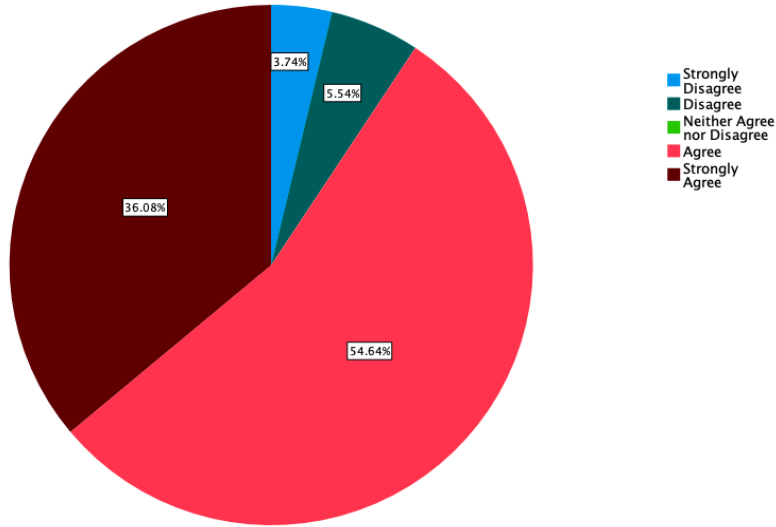

**Figure 2.** Frequency of responses to DaD Control Scale Item 2: *Users should have the option to configure how their account should be managed by a service provider after they pass away.*

Figure 2 indicates that the majority of respondents Agree or Strongly Agree that they should have an option to configure how their data is managed after they pass on. However, as indicated in Figure 3 below, the majority of respondents have not developed a will that covers their digital presence.

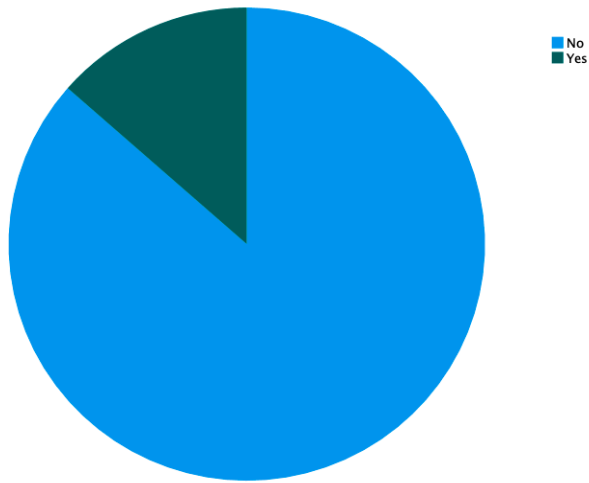

**Figure 3.** Frequency of responses to item: *"Do you have a 'digital will' or clause in your existing will that deals with your digital presence after you pass away?"*

**Who do Australians trust to manage their data after death?**
To explore the second research question, we examined the participants' responses to the following items: *If you were able to create clear instructions to be implemented on your data after you pass away, who would you <u>trust</u> to implement these instructions?* And *If you were able to create clear instructions to be implemented on your data after your pass away, which of the following options would you find more <u>convenient</u> to use?*

As indicated in Figures 4 and 5, participants preferred to entrust a close friend or family member with managing their data after death. Failing this, they preferred to use a third-party application that they could configure themselves. Both options outweighed the least preferred option: to rely on the settings provided by the online vendor, which was ranked the lowest in terms of both trust and convenience.

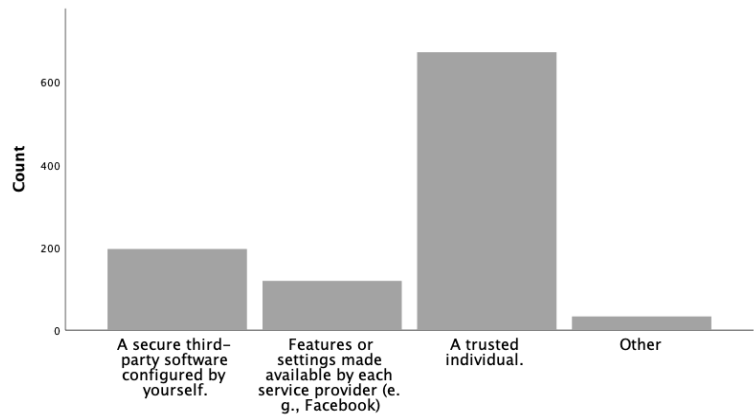

**Figure 4.** Participant responses to item: *If you were able to create clear instructions to be implemented on your data after you pass away, who would you **<u>trust</u>** to implement these instructions?*

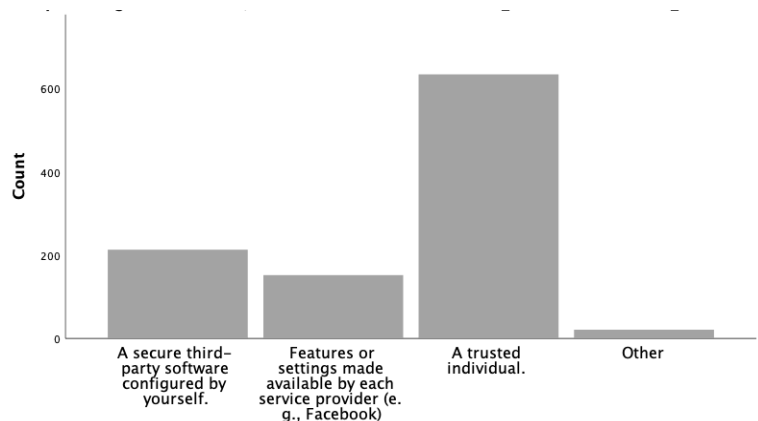

**Figure 5.** Participant responses to item: *If you were able to create clear instructions to be implemented on your data after your pass away, which of the following options would you find more **<u>convenient</u>** to use?*

**Do these preferences and trust levels change by demographic?**
Table 3 reports the result of an exploratory regression predicting scores on the DaD Control Preference Scale. Unexpectedly, Age and Gender did not predict preferences for data management after death, nor did marital status. However, internet usage, education, and the number of children did predict preferences for data after death.

**Table 3.** A linear regression of demographic variables predicting scores on the DaD Control Preference Scale.

| Independent Variable | t |
|---|---|
| Age | 0.40 |
| Gender | 1.21 |
| Internet Usage | 2.67** |
| Education | 2.55* |
| Marital Status | -1.28 |
| No. Children or Dependents | 5.23*** |

*$p<.05, **p<.01, ***p<.001$*

**RQ4) What should any new solution provide to meet the needs of Australians wishing to secure their data after death?**

A plurality of participants (34.0%, n=347) indicated that they wished that some of their data should be shared with their family and friends but not all. Almost as many participants (30.1%, n=307) wished for their data to be entirely deleted, while a smaller subset wished for the data to be edited before release (10.6%, n=108). Figure 6 provides an overview of the count data. Note participants could select more than one option.

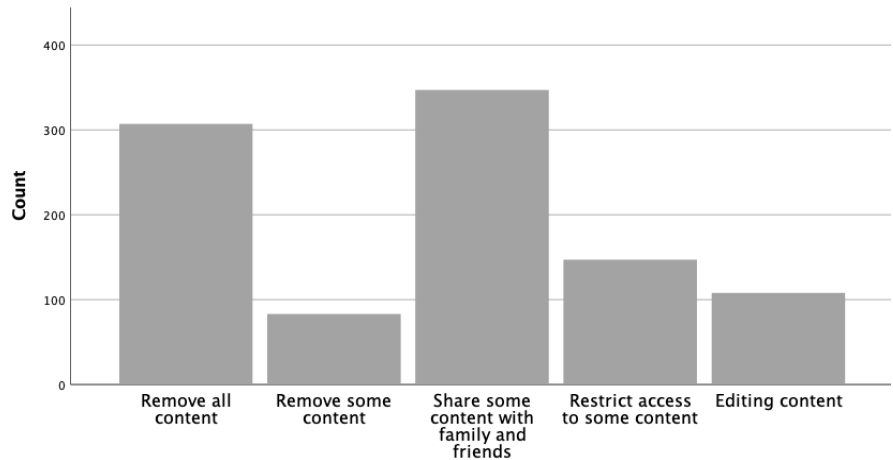

**Figure 6.** Counts of responses to item: *What type of instructions would you like to apply to the data in your online accounts after you pass away?*

## 3     Discussion

The current study surveyed 1020 Australians to explore their preferences for the management of their data after their death. As expected, most Australians in our sample wanted some level of control over how their data is managed after death and a plurality indicated that they would prefer full access to all accounts be given to one trusted individual to manage. However, the majority did not already have a will that covered the management of their digital presence. Participants preferred to entrust the management of their data to a trusted close individual or a third-party software that they could administrate themselves. Social media companies were ranked low regarding both trust and convenience to manage data after death. Australians who were highly active on the internet users had stronger need for some level of control over the data after death. People with children or with greater levels of education were also more likely to express a desire to have control over the data after death. Unexpectedly, marital status, age, and gender did not predict data preferences. Regarding the development of a technological solution, participants indicated that they would like their data to be mostly removed, but that certain content should be preserved and shared with family and friends.

As expected, demographic differences did show some relationship with posthumous data management preferences. However, while having children was one such factor, marital status was not. This may indicate that Australian's are concerned regarding the legacy of their online presence more so in relation to their progeny than their spouse. Understandably, Australians with children may wish to ensure that certain elements of their online identify is preserved for future generations. It also appears that the more educated Australians and those who spend a greater amount of their lives on the internet are most likely to have strong preferences for the management of their online life post-mortem.

While most Australians prefer to entrust the management of their post-mortem data to a trusted individual, it is likely that the efficient operationalisation of this would require a third-party software solution. That is, for an individual to take control of all accounts at the point of death would require a great degree of work to reset accounts and gain access, particularly given the breadth of different approaches taken by vendors to address this issue (Sikun, 2022). Therefore, it is telling that the second most preferred option by Australians was to entrust the management of their data after death to a secure third-party software solution that they can configure while alive. Such a solution could interface with the major online vendors (social media, cloud hosting etc.) to action the deceased's will – deleting some data, while sharing other data with selected individuals. Prototypes of such technologies have been produced, and even brought to market, but often fail to produce a sustainable business model (Nansen et al., 2023)

Given the observed distrust of social media, both here and in previous studies (e.g., Hsieh-Yee, 2021; Liu, 2015; Radu et al., 2016; Rajković et al., 2021; Sikun, 2022; Tang & Liu, 2015) and the potential risks associated with posthumous data management, it becomes imperative for the academic and tech communities to direct focus to this issue. Understanding the desires and apprehensions of the general populace regarding posthumous data management is more crucial now than ever. The aim should be to conceptualise and develop a third-party solution that circumvents these prevalent issues. Such a solution would not only address the immediate concerns of data management after death but also sidestep the looming shadow of mistrust that currently taints engagements with the large social media entities.

## 6   Conclusion

Managed appropriately, the data generated by today's online communities can form valuable keepsakes for future societies. Relatives, descendants, and researchers alike will appreciate the unprecedented insight into the lives of long-past cultures. This study sought to uncover the preferences of Australians regarding the management of their own data after death and to further the conversation regarding ways to balance the value of this data to future society with the privacy needs of the individual. We found that most Australians desired a level of control over how their data is managed after death, and they preferred to entrust the management of their data to a trusted close individual or a third-party software that they could administrate themselves. Social media companies were ranked low regarding both trust and convenience to manage data after death. Furthermore, we found that highly active internet users are more likely to have stronger preferences for some level of control over their data after death, as did people with children and with greater levels of education. Unexpectedly, marital status, age, and gender did not predict preferences for control of data after death. Future re-search focus should be to conceptualise and develop a third-party solution that enables these preferences to be realised. Such a solution could interface with the major online vendors (social media, cloud hosting etc.) to action the deceased's will – deleting some data, while sharing other data with selected individuals.

**Acknowledgments.** Dr Arash Shaghaghi (project lead) acknowledges the partial funding from Deakin Cyber Research and Innovation Centre.

**Disclosure of Interests.** The authors have no competing interests to declare that are relevant to the content of this article.